\documentclass[final,5p,times,twocolumn]{elsarticle}   
\usepackage{lineno,hyperref}
\usepackage{float}
\usepackage{subfig}
\journal{Nuclear Instruments and Methods in Physics Research, Section A, }

\begin{document}

\begin{frontmatter}
	
\title{LUXE-PROC-2022-013\raggedleft\\R\&D for compact 
calorimeters for $e^+e^-$ colliders and their application to the LUXE experiment\tnoteref{n1}}

\tnotetext[n1]{on behalf of \textbf{FCAL} and \textbf{LUXE} Collaborations}

\author[iss]{Veta Ghenescu\corref{cor1}}
\cortext[cor1]{Corresponding author}
\ead{veta.ghenescu@cern.ch}	
\address[iss]{Institute of Space Science, Atomistilor 409, P.O. Box MG-23, Bucharest-Magurele RO-077125, ROMANIA}
\begin{abstract}
The\textbf{ FCAL} collaboration is preparing large-scale prototypes of special calorimeters to be used in the very forward region at future electron-positron colliders. Two compact calorimeters are foreseen, LumiCal and BeamCal, designed as sandwich calorimeters with very thin sensor planes to keep the Molière radius small. Silicon sensor prototypes and dedicated FE ASICs have been developed and produced for LumiCal. The ASICs have been designed to cope with the timing and dynamic range requirements. 
The electromagnetic calorimeters of the \textbf{LUXE} experiment are based on a similar technology to that developed by the \textbf {FCAL} collaboration.  \textbf{LUXE} will measure the number of positrons and their energy spectrum in e-laser and $\gamma$-laser interactions or electrons in $\gamma$-laser interactions. Results of recent beam tests will be presented.
\end{abstract}
\end{frontmatter}
\section{Introduction}
The instrumentation of the very forward region \cite{intro_1, intro_2} 
at a future ${e}^+{e}^-$ collider is facing numerous challenges: the angular coverage down to the smallest polar angles requires good radiation hardness of the sensors for the forward calorimeters, while reconstruction of the EM shower to be associated with a high-energy electron is performed in the presence of intense background. 
Two dedicated calorimeters \cite{intro_3} are foreseen of high energy and polar angle resolution to instrument the very forward region of the future ${e}^+{e}^-$ collider detectors, called LumiCal and BeamCal. The position and size of the apertures of the calorimeters are dictated by beam-induced backgrounds at low angles. The same background drives radiation hardness requirements, as well as the trade-off between granularity and occupancies. 
In recent times, there have been numerous studies of particle physics in intense electromagnetic fields and nonlinear QED. 
\begin{figure}[!htb]
\centering
\includegraphics[width=1.0\linewidth, height=0.79\linewidth]{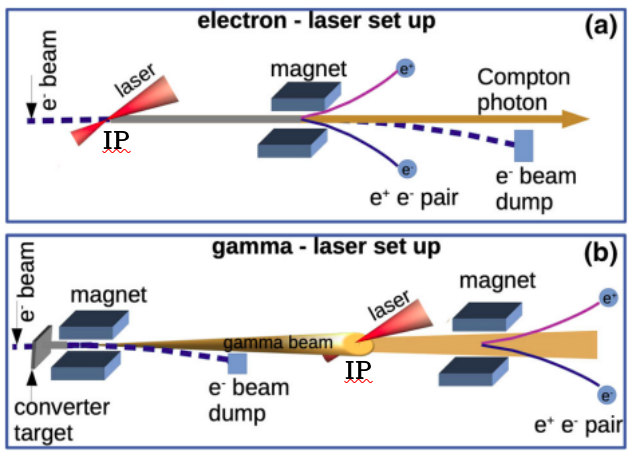}
  \caption{Schematic experimental layouts for the \textbf{(a)} e-laser and \textbf{(b)} $\gamma$-laser set-up.}\label{fig:luxe_a_b}
\end{figure}
    The LUXE experiment \cite{intro_4} proposed at DESY and the European XFEL (Eu.XFEL) in Hamburg is intended to study strong-field QED processes using collisions between a high power optical laser and the 16.5 GeV electron beam of the Eu.XFEL, as well as collisions of the laser with high-energy secondary photons.  Due to the nature of the production processes, the signal expected across each system varies over many orders of magnitude, from ${10}^{-4}$ to ${10}^9$ per 1 Hz laser bunch crossing (BX), depending on operating mode and particle type. Schematic layouts of the LUXE experiment are shown in Figure~\ref{fig:luxe_a_b} for the two configurations envisaged for the e-laser and the $\gamma$-laser set-ups. In both cases a strong field will be presented at the interaction point (IP). For the e-laser set-up the electrons are directly guided to the IP, where a laser beam is directed at the same time. The electrons and positrons produced in these collisions are deflected by a magnet and then detected in a variety of detectors optimised for the expected fluxes of particles: positrons are detected by a silicon pixel tracking detector and a high-granularity calorimeter, while electrons are measured by a scintillation screen and gas Cherenkov detectors. Photons that are produced at the IP continue along the beamline towards a photon detection system that is designed to measure their flux, energy and spatial distribution.
The $\gamma$-laser set-up is similar, except that either a target is placed into the beam to produce a broadband beam of bremsstrahlung photons or a low-power laser is used to create a monochromatic photon beam via inverse Compton scattering, and the technology for the electron side is adapted due to the lower particle rates expected. Three signature processes of QED are to be measured at LUXE: non-linear Compton scattering; non-linear Breit-Wheeler pair production and the non-linear trident process. 
%
\section{\textbf{LumiCal} prototype performance}
\begin{figure}[!htb]
\centering
  \includegraphics[height=4.5cm,keepaspectratio]{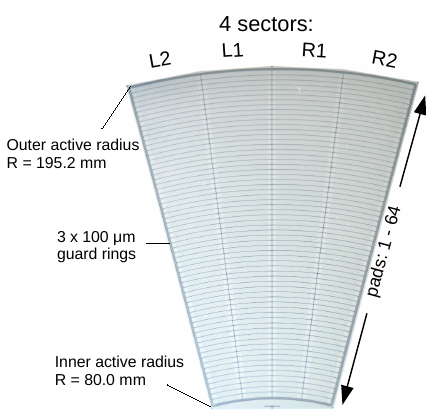}
  \caption{Si sensor segmentation.}\label{si_sensor}
\end{figure}

\begin{figure}[!htb]
\centering
  \includegraphics[height=3.5cm,keepaspectratio]{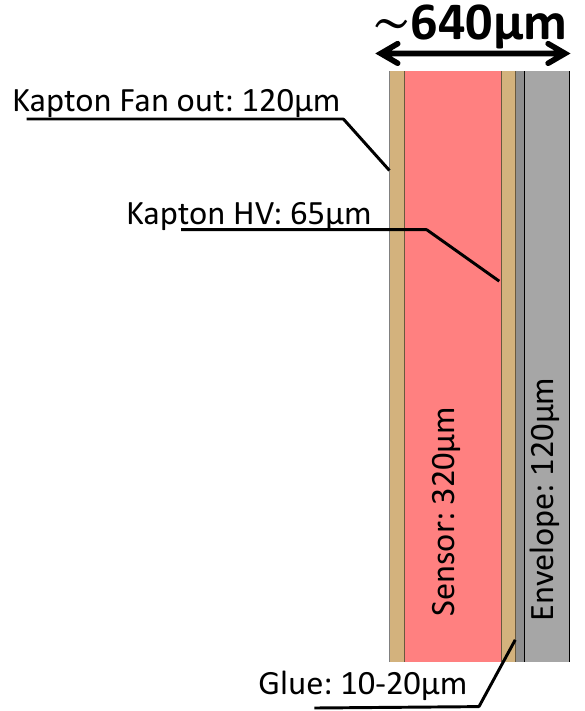}
  \caption{Structure of the LumiCal detector plane.}\label{detector_module}
\end{figure}

\begin{figure}[!htb]
\centering
  \includegraphics[height=5.5cm,keepaspectratio]{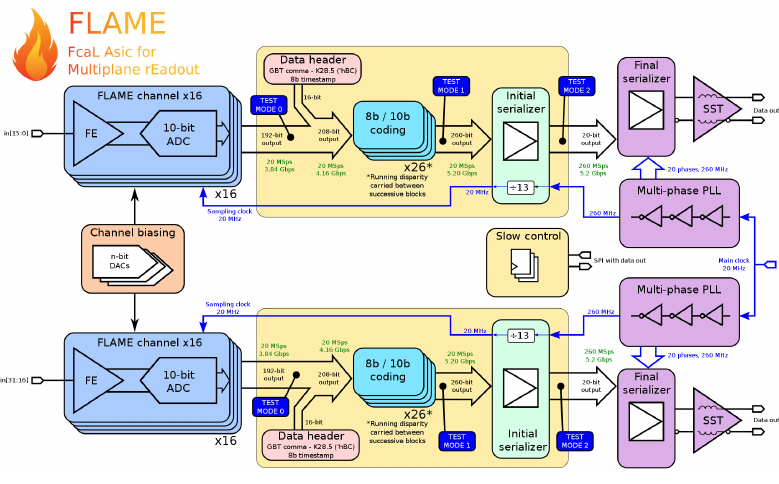}
  \caption{Block diagram of a 32-channels FLAME ASIC arhitecture.}\label{flame}
\end{figure}

\begin{figure}[!tbh]
\centering
  \includegraphics[height=5.5cm,keepaspectratio]{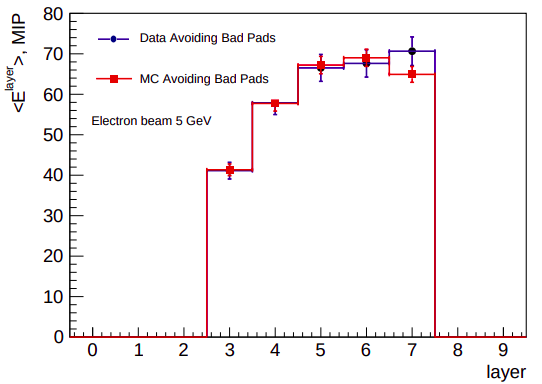}
   \caption{Longitudinal shower profile, comparison between data and MC simulation. The distributions are obtained with a 5 GeV electron beam.}\label{lon_shower}
\end{figure}

\begin{figure}[!htb]
\centering
  \includegraphics[height=6.8cm,keepaspectratio]{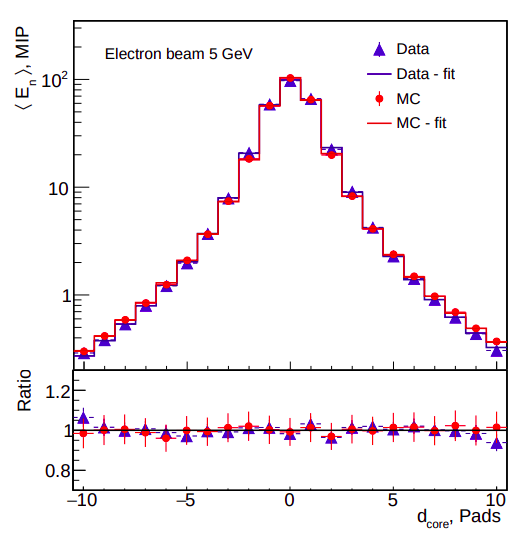}
  \caption{The transverse shower profile E\textsubscript{m}, 
as a function of d\textsubscript{core} in units of pads,
from beam-test data and the MC simulation,
after symmetry corrections and fit. The lower
part shows the ratio of the distributions to the fitted function, for the data
(blue) and the MC (red).}\label{transvers_sh}
\end{figure}

\begin{figure}
   \centering
  \includegraphics[height=5.5cm,keepaspectratio]{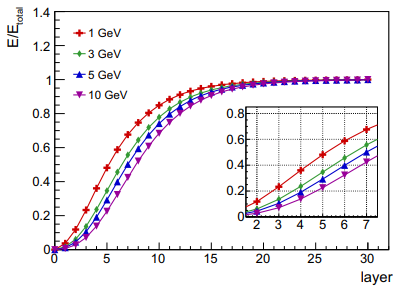}
  \caption{Cumulative distribution of the fraction of energy deposited in
the detector planes as a function of the number of layers for different
electron beam energies. The insert shows an expanded view of the
region for planes 2 to 7.}\label{shower_contain}
\end{figure}

LumiCal is primarily designed for high luminousity measurements. In the current prototype, 
tungsten absorber plates of one radiation length thickness are interspersed with silicon detector planes. These n-type silicon sensors are 320$\mu$m thick, shaped as a ring segment of 30 degrees, subdivided into four sectors of 7.5 degrees each. In a radial direction there are 64 pads in each sector with a pitch of 1.8 mm as is illustrated in Figure~\ref{si_sensor}. 
Two thin flexible Kapton-cooper foils are used to supply high voltage and to connect the read-out electronics board to the sensor. Figure~\ref{detector_module} shows a detector plane of about 640$\mu$m thickness which was installed in a 1mm gap. Until the 2020 test-beam campaign the FCAL Collaboration used the APV-25 ~\cite{apv25} chip hybrid board as a temporary solution. 
For the last test-beam new readout boards based on a new readout ASIC called FLAME~\cite{flame} (FcaL Asic for Multiplane rEadout) 
were manufactured and were connected to the LumiCal detectors to record the channel signals. FLAME was designed in 130 nm CMOS technology and matches the requirements of future colliders; in particular it has a larger input signal dynamic range compared to the APV-25 chip. It consists of a pair of two identical 16-channel blocks as is shown in Figure~\ref{flame}. 
The main goals of the test-beam campaigns were to measure the electromagnetic shower development in longitudinal and transverse directions and the effective Molière radius. The experimental results were compared with prediction of GEANT4 Monte Carlo (MC) simulations ~\cite{geant_MC} where the experimental setup was implemented. 
The longitudinal development of electron showers is shown in Figure~\ref{lon_shower} in terms of average
shower energy deposits per detector plane as a function of the number of layers. 
The average energy $\langle{E}_{l}^{\rm layer}\rangle$ deposited in calorimeter layer \textit{l}
is calculated as the following:
\begin{equation}
$$\langle{E}_{l}^{\rm layer}\rangle = \sum_{n}\langle{E}_\textit{nl}^{\rm det}\rangle$$
\end{equation}
where \textit{n} runs over the radial pads of the two central sectors
of the sensor. About 5\% of randomly distributed channels
in the calorimeter have a larger noise level corresponding to
signal sizes of up to 40 MIPs.
The development of the longitudinal shower profile is measured using only events with properly working channels.
In Figure~\ref{lon_shower} the deposited energy as a function of the layer \textit{l}
is shown for data and Monte Carlo simulation. The maximum of the shower is reached in data at layer 7. Both distributions are, within statistical uncertainties, in reasonable agreement.
The segmentation of the LumiCal detector plane is rather specific and a method was developed ~\cite{shower_1} to take into account its circular shape to measure the Molière radius. An example of the average transverse shower profiles at 5 GeV beam energy is
shown in Figure~\ref{transvers_sh}. The data are again well described by the results of simulations. 
The transverse shower profile (Figure~\ref{transvers_sh}) is fitted with a function $F_{E}$ (Eq. \ref{eq:2}), where a Gaussian term describes a shower core and a modified Grindhammer-Peters term ~\cite{grindhammer} describes the tail of the profile:
\begin{equation}
\label{eq:2}
 \textit{F}\textsubscript{~\textit{E}}(r) = A_{C}e^{-(\frac{r}{R_{C}})^{2}} + A_{T}\frac{2r^{\alpha}\mathrm{R}_{T}^{2}}{(r^{2} + \mathrm{R}_{T}^{2})^{2}},
\end{equation}
where $A_{C}$, $R_{C}$, $A_{T}$ , $R_{T}$ and $\alpha$ are free parameters to be determined by fitting $F_{E}$ to the measured distribution. Integration over ${r}$, from 0 to $R_{M}$, defines 90\% of the deposited energy. The effective Moliere radius $R_{M}$ is found to be: 8.1 ${\pm}$ 0.1(\textit{stat}) ${\pm}$ 0.3(\textit{syst}) \textit{mm} ~\cite{shower_2} for 5 GeV electrons.
The effective Molière radius depends a bit on the electron energy due to a limited longitudinal coverage
with the existing number of detector planes. From simulation, it is clear that 20 planes should
suffice even for 10 GeV electron shower containment as shown in Figure~\ref{shower_contain}.
The most recent test beam 2020 campaign was dedicated to studies of the performance of deep LumiCal prototype with 15 detector planes. 
\section{Particle detectors in LUXE experiment}
\begin{figure}[!htb]
\centering
 \includegraphics[height=3.055cm,keepaspectratio]{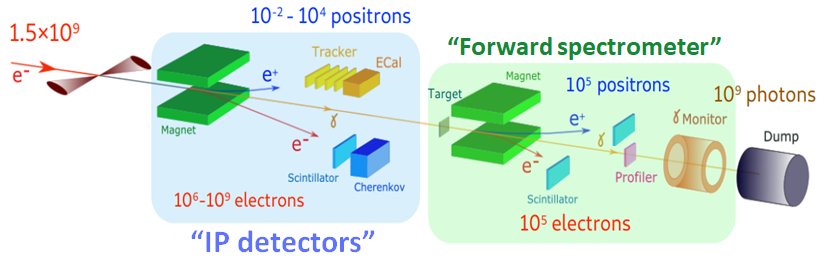}
  \caption{ Diagram of the LUXE experiment layout. Shaded areas indicate electron positron spectrometer
(blue) and photon spectrometer (green).
}\label{pds_sys}
\end{figure}

\begin{figure}[!htb]
\centering
  \includegraphics[height=6.05cm,keepaspectratio]{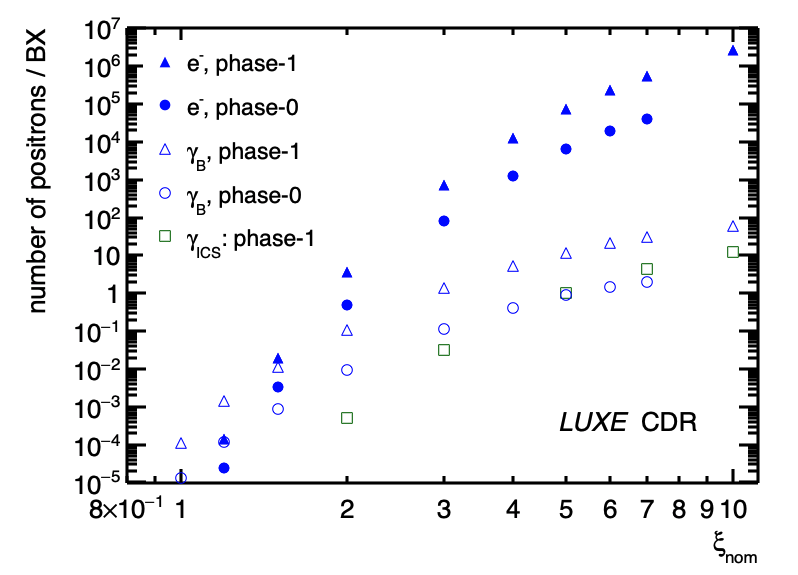}
  \caption{Number of positrons per bunch crossing produced in the e-laser and $\gamma$-laser
setups for phase-0 and phase-1, as a function of $\xi$\textsubscript{nom}. The electron beam energy is set to
16.5 GeV, and the laser waist parameter varies between 100 $\mu$m and 3 $\mu$m in the range of
$\xi$.}\label{pos_rate}
\end{figure}

\begin{figure}[!htb]
\centering
  \includegraphics[height=4.25cm,keepaspectratio]{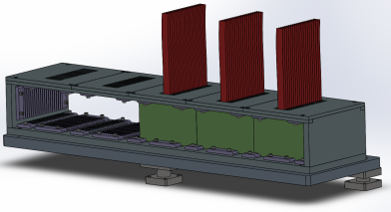}
  \caption{A sketch of the mechanical frame of the ECAL calorimeter. The frame holds the tungsten absorber plates, interspersed with assembled 
detector planes (in green). The front-end electronics will be positioned on top of ECAL (brown).}\label{ecal_fig}
\end{figure}

\begin{figure}[!htb]
\centering
\includegraphics[width=1.0\linewidth, height=0.65\linewidth]{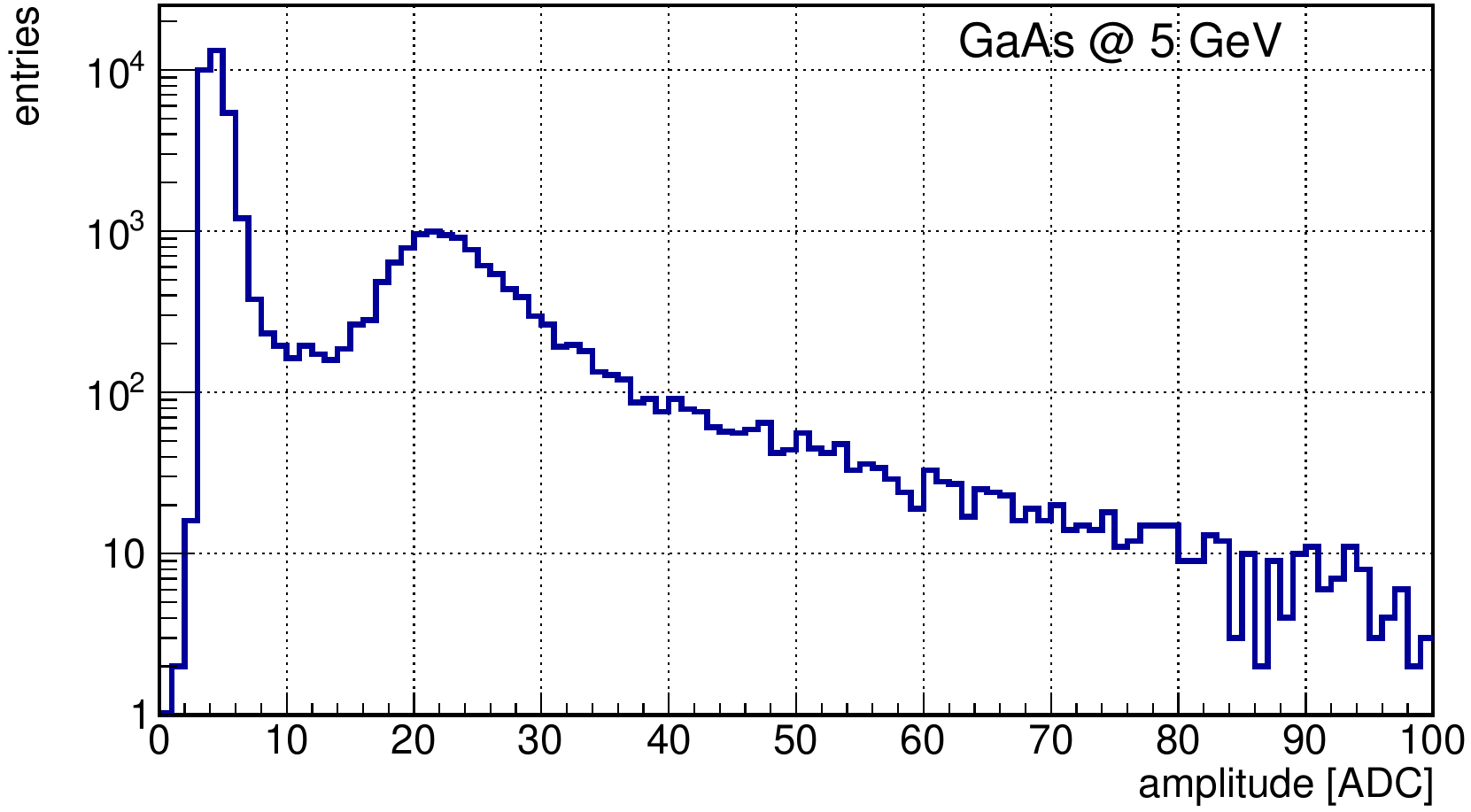}
\caption{Signal distribution for all GaAs sensor channels.}\label{sig_gaas}
\end{figure}

\begin{figure}[!htb]
\centering
 \includegraphics[width=1.0\linewidth, height=0.6\linewidth]{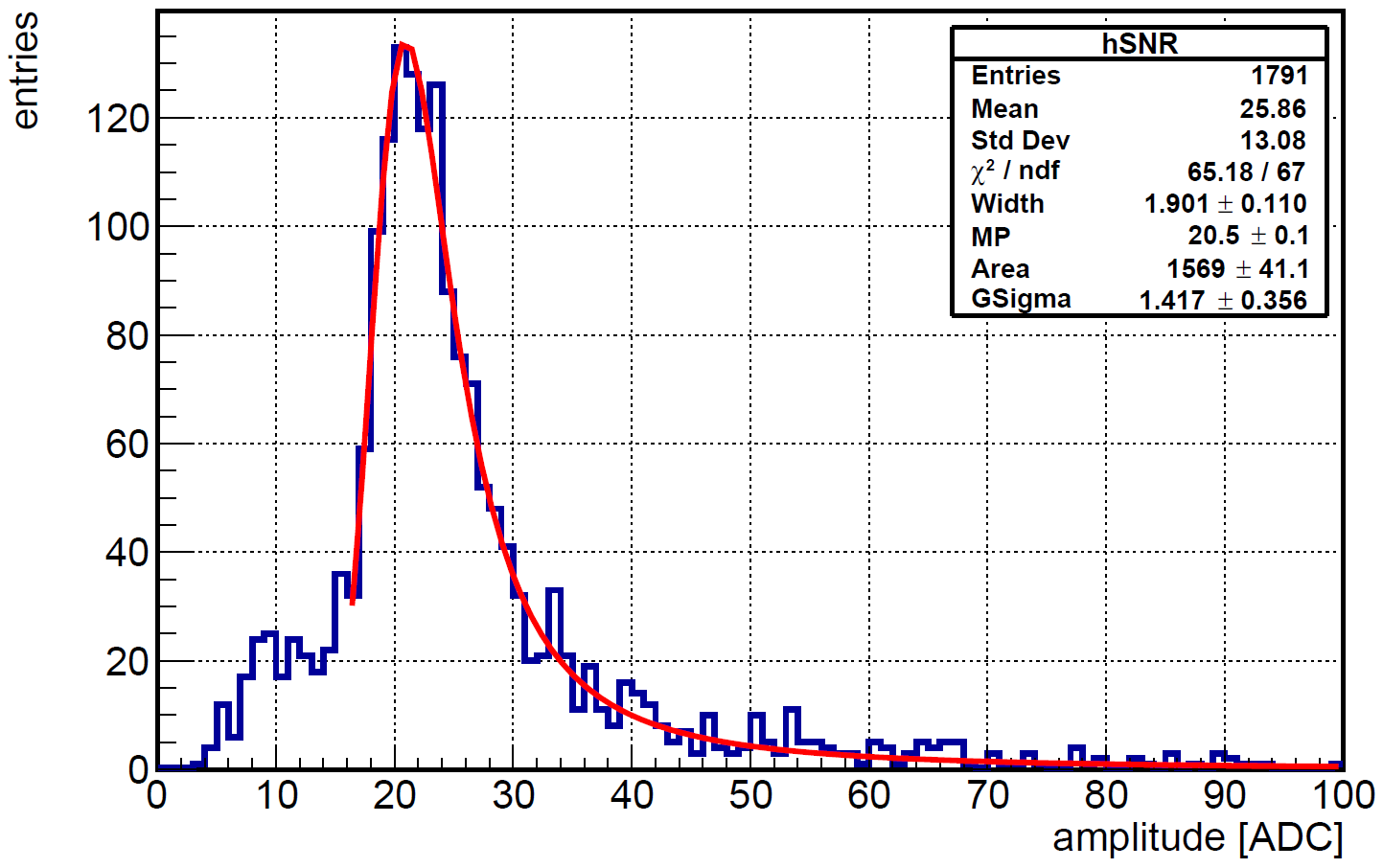}
  \caption{Signal distribution in a single channel of GaAs sensor fitted with LG convolution function.}\label{sig_dist_gaas}
\end{figure}

The technologies developed for LumiCal prototype are integrated in the Positron Detection System (PDS) of the LUXE experiment. The PDS consists of a silicon pixel tracker and a sensor-tungsten sandwich calorimeter (ECAL). The ECAL will allow to measure independently of the tracker the number of positrons, and their energy spectrum. The LUXE experimental setup for the electron-laser mode is shown in Figure ~\ref{pds_sys}. 
For phase-0, at the start of the experiment, a laser power of 40 TW is assumed, while in phase-1 the laser power is increased to 350 TW. The positron yield per bunch crossing for e-laser and $\gamma$-laser interactions is shown in Figure~\ref{pos_rate} as function of $\xi$, where $\xi$ is the intensity parameter of the laser field. 
Different values of $\xi$ are achieved by varying the laser power. It is seen that for the e-laser mode the positron yield is larger than for the $\gamma$-laser one by a factor of $\sim $ 10–10000,
depending on $\xi$. For phase-0 of the e-laser run, up to 10\textsuperscript{5} positrons are expected per bunch crossing, while for the $\gamma$-laser running at low $\xi$, most of the bunch crossings are without a positron, and only $\sim $ 10\textsuperscript{2} are expected at the highest $\xi$. For phase-1 the rates are roughly 10–100 times larger.
Close to the interaction point, the trajectory of the produced electron positron pair is bent by a dipole magnet and on the left side of the beam direction, a tracker and an electromagnetic calorimeter (ECAL) are installed for the positron detection. Essential for the performance of ECAL are high granularity for very good position resolution, compactness, i.e.
a small Molière radius, to ensure a high spatial resolution of local energy deposits, and good energy resolution
to measure and infer the spectrum of positrons. The ECAL is designed as a sampling calorimeter composed of 20 layers of 3.5 mm thick tungsten absorber plates and assembled sensor planes placed in a 1 mm gap between absorber plates as shown in Figure~\ref{ecal_fig}. Two types of sensors are under study, silicon and gallium arsenide. 
These sensors were measured in November 2021 at the DESY-II Synchrotron using electrons with energies between 1 to 5 GeV. The preliminary results of the signal distribution in GaAs sensor is shown in Figure~\ref{sig_gaas}. As can be seen a clear signal of minimum ionising particles is visible, well separated from the noise. The size of the signal matches the expectations from the energy deposited by the relativistic electrons in the sensor material. 
An example of the typical signal distribution produced by 5 GeV electron beam and measured in a single channel of the GaAs sensor is shown in Figure~\ref{sig_dist_gaas}. 
The most probable value (MPV) of the peak is estimated using a fit with a Landau–Gauss convolution function. The data analysis is in progress.
\section{Conclusions}
Major components developed by the FCAL Collaboration can be operated in future colliders and are considered for the LUXE experiment. LUXE will be one of the first experiments to explore QED in the uncharted strong-field frontier. The experimental setup that can be adapted to a large dynamic range has been designed and is going to be tested and installed in the distribution fan of the XFEL.EU located in Osdorfer Born. The first data taking of LUXE is scheduled for the 2026.
\section*{Acknowledgments}
This work was partially supported by the Romanian Ministry of Research,
Innovation, and Digitisation, grant no. 16N/2019 within the National Nucleus Program. The measurements leading to these results have been performed at the Test Beam Facility at DESY Hamburg (Germany), a member of the Helmholtz Association (HGF).

\end{document}